\documentclass{emulateapj}

\newcommand{\etal}{{et al.}}

\shorttitle{Steady GRMHD Inflow/Outflow Solution}
\shortauthors{H.-Y. Pu et al.}

\begin{document}
\title{Steady General Relativistic Magnetohydrodynamic Inflow/Outflow Solution along Large-Scale Magnetic Fields that Thread a Rotating Black Hole}

\author{Hung-Yi Pu\altaffilmark{1}, Masanori Nakamura\altaffilmark{1}, Kouichi Hirotani\altaffilmark{1}, Yosuke Mizuno\altaffilmark{2,3}, Kinwah Wu\altaffilmark{4}, and Keiichi Asada\altaffilmark{1} }
\affil{
$^1$Institute of Astronomy \& Astrophysics, Academia Sinica, 11F of
Astronomy-Mathematics Building, AS/NTU No. 1, Taipei 10617, Taiwan}
\affil{$^2$Institute of Astronomy, National Tsing Hua University, Hsinchu 30013, Taiwan}
\affil{$^3$Institute for Theoretical Physics, Goethe University, D-60438, Frankfurt
am Main, Germany}
\affil{$^4$Mullard Space Science Laboratory, University of College London, Holmbury St Mary, Dorking, Surrey RH5 6NT, UK \\
}

\begin{abstract}
General relativistic
magnetohydrodynamic (GRMHD) flows along magnetic fields threading a black hole can be divided into inflow and
outflow parts,  according to the result of the  competition between the black hole gravity and
magneto-centrifugal forces along the field line. 
Here we present the first self-consistent, semi-analytical solution for 
a cold, Poynting flux-dominated (PFD) GRMHD flow, which passes all four critical (inner and outer,
Alfv\'en and fast magnetosonic) points along a parabolic streamline.
By assuming that the dominating (electromagnetic) component  of the energy flux per flux tube
is conserved at the surface where the inflow and outflow are separated, the outflow part of the solution can be constrained by the inflow part.
The semi-analytical method can provide fiducial and complementary solutions for GRMHD
simulations  around the rotating black hole, 
given that the black hole spin, global streamline, and magnetizaion (i.e., a
mass loading at the inflow/outflow separation) are prescribed. 
For reference, we demonstrate a self-consistent result with  the work by McKinney in a quantitative level.

\end{abstract}

\keywords{galaxies: active-----
galaxies: jets --- Magnetohydrodynamics (MHD) --- Black hole physics}


\section{Introduction}
\label{sec:int}
Relativistic jets emerging from accreting black hole systems have been
observed in active galactic nuclei (AGN), micro-quasars (stellar mass
black hole X-ray binaries), and presumably gamma-ray
bursts (GRBs). Observationally, 
the bulk Lorentz factors $\Gamma$ of jets in AGNs are $\gtrsim 10 - 20$
\citep{J05,Co07,Gu09,P09,L13}, and the values could be higher for some
blazars \citep[see][]{GK06,HV09}.  The Lorentz factors of micro-quasar
jets are lower, mostly with $\Gamma
\sim 2 - 10$ \citep[e.g.][]{Fd04, C02}, but still there are a few found
to have $\Gamma> 10$ \citep[see][]{MJ06}. 
Jets in gamma-ray bursters are
supposed to be ultra-relativistic,
and their Lorentz factors can
be as high as $\sim 100 - 1000$ \citep[see,
e.g.][]{LYB10,Lyu11}.
How jets become relativistic after being
launched from nearby black holes is a
long-standing issue. 
Electromagnetic or magnetohydrodynamic (MHD) mechanisms are
frequently invoked to extract energy and momentum from the black hole
and accretion disk \citep[e.g.,][for reviews]{MKU01}. One of the key
issues to be addressed is the potential of the MHD flow acceleration up
to a high bulk Lorentz factor of $\Gamma >10$.

An ideal engine to power 
relativistic jets is a
spinning black hole.
Close to
the black hole, the rapid winding of the azimuthal
component in large-scale magnetic fields due to the frame-dragging
inside the black hole ergosphere
results in a counter
torque (induced by the Lorentz force) against a black hole rotation. The
energy that the black hole spent to perturb the field line can be
propagated outward in the form of torsional Alfv\'en
waves, thus extracting the black hole energy electromagnetically
 \citep[][hereafter
BZ77]{BZ77}. However, because the environment around an accreting black hole is not a
perfect vacuum (contrary to the force-free treatment in
BZ77), the general relativistic magnetohydrodynamics
(GRMHD), which consist of electromagnetic and fluid components,
provide a more general picture for the dynamics and structures
 of relativistic jets in both theoretical approaches
\citep[e.g.][]{C86a,C86b,C87,T90,FG01,F96,F04} and numerical simulations
\cite[e.g.][]{K98, K00, M04,MG04, HK06, M06, B08, T10, T11, TM12}.

\begin{figure}
\begin{center}
\includegraphics[scale=0.4, trim=0.6cm 0.2cm 0.2cm 0.2cm, clip=true]{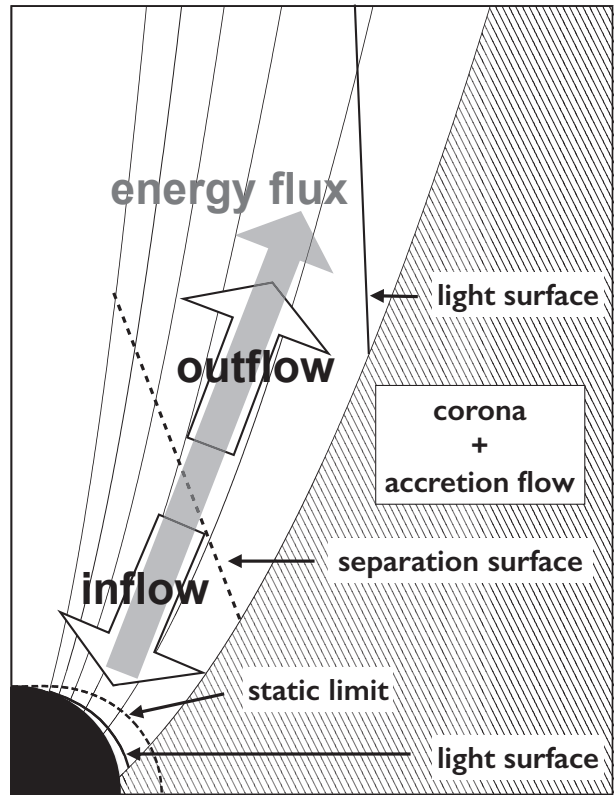}
\caption{
Schematic illustration of a Poynting flux-dominated (PFD)  GRMHD flow confined by the accretion flow and its corona.  
The outward-streaming curves indicates  ordered, large-scale magnetic fields
  that thread the black hole event horizon. 
The inflows and the outflows (represented by thick white arrows) are along the field lines,  
  and they are separated by the separation surface (marked by a dashed line). 
The energy flux (represented by a gray arrow) is outward in both the inflow and outflow regions, 
  as the black hole rotational energy is extracted and transported outward.
The static limit (dashed curve) and the light surface (solid curve) 
 outside the  black hole (black region) are also shown.   
 }
\label{fig:sketch}
\end{center}
\end{figure}

 The overall  configuration of  an accreting black hole system is schematically illustrated in Figure
\ref{fig:sketch} \cite[see also GRMHD simulations for magnetized accretion, e.g. ][]{M06, MG04, HK06}. Ordered, parabolic lines are developed near the funnel,  which  is confined by the corona and/or accretion flow.
Due to the relative absence of accreting materials, the funnel region is Poynting flux-dominated (PFD). The fluid loading onto 
the field is accelerated inward (or outward) if  the black hole
gravity force is larger (or smaller) than the ``magneto-centrifugal'' forces \citep[e.g.][in the case
of the accretion disk]{sad10}.
As  pointed out in the theoretical work  of \citet{T90}, black hole rotational energy 
can be extracted outward by a PFD GRMHD inflow. This is a direct result of that the electromagnetic components dominating the GRMHD flow, and the electromagnetic component is responsible for extracting the black hole energy, similar to the BZ77  process.
The outward energy flux, after being extracted from the black hole, is expected to propagate  continuously  outward throughout the magnetosphere from
the inflow region to the outflow one. 
In this paper, we focus on the PFD GRMHD flow in the funnel region, including both the inflow and outflow parts.

For comparison, let us quickly consider the case when the GRMHD flow becomes fluid-dominated.
In that case, the energy flux is dominated by the fluid component, and therefore
it has an inward direction for inflow, but outward for
outflow (c.f., the energy flux direction shown in Figure \ref{fig:sketch}). 
Such discontinuity of the energy and momentum fluxes implies that the
outflow is accretion-powered, which is constrained by the
energy input from the disk/corona. 
The switch-on and switch-off of the extraction of the black hole energy (inflow) may closely relate to 
the launching and quenching of relativistic jets (outflow) \citep[e.g.][]{pu12,glo13}.

Prior to the GMRHD studies mentioned, \citet[]{Phi83}
considered the inflow and outflow along a monopole field jointly by the conservation of the total energy flux per flux tube.
In this pioneering work, they consider energy extraction
from the black hole via BZ77 process (the inflow part),  and the Michel's ``minimum torque
solution'' \citep[]{Mic69}, in which the fast(-magnetosonic) point is located at infinity (the outflow part). 
We, however, suggest that a more realistic situation can be considered: the black hole energy extraction  process in the framework of GRMHD, and  a type of parabolic GRMHD flows as a result of external pressure confinements provided by the corona/accretion. Recent  observational evidence also supports this idea; nearby active radio galaxy, M87, exhibits the parabolic streamline up to $\sim 10^{5}$ Schwarzschild radius \citep[]{AN12}.

Furthermore, we are interested in the case that the fast point of the outflow  is  located at a  finite distance.
This consideration is directly related the conversion
from Poynting to kinetic energy fluxes of the flow and therefore the jet acceleration.  
Poloidal magnetic
flux is required to diverge sufficiently rapidly in order for most of
the Poynting flux  to be
converted into the kinetic energy flux beyond the fast
point (also known as the magnetic nozzle effect \cite[e.g.][]{C89,LCB92,BL94,T98} .

\citet[][]{BN06} examine the acceleration of the jet along a parabolic streamline by introducing
a small perturbation into the force-free field. As a result, the fast point is located at a finite distance.
This indicates how plasma loading in the flow plays a role in accelerating the flow, as well as a conversion from Poynting
to kinetic/particle energies.
They consider the behavior of the outflow in the flat spacetime. However, we are  interested in both the inflow and outflow near a black hole.

All of these theoretical works provide important pieces toward a   picture that includes the following process along the  field line: (i) in the inflow region  the rotational energy of the black hole is extracted outward by the GRMHD inflow, (ii) at the the inflow/outflow separation surface the extracted energy flux is carried out continuously, and (iii) in the outflow region  the flow passes the fast point, and hence the bulk Lorentz factor increases. Although this picture has been already recognised  in the quasi-steady state in GRMHD simulations \citep[e.g. ][]{M06, MG04, HK06}, no steady solution is available in the literature. 

In this paper, we present the first semi-analytical work. We consider the energy extraction
from the black hole via the GRMHD (inflow), and the perturbed force-free parabolic field line in \citet[]{BN06} (outflow). 
With given black hole spin, field angular velocity,  and magnetization at the separation surface, we are able to to constrain the outflow solution by the inflow solution. For a reference, we adopt similar parameters reported in the GRMHD simulation of \citet[][;hereafter M06]{M06}. Our semi-analytical solution passes all the critical points (inner and outer,
Alfv\'en and fast points), and agrees with the  inflow and outflow properties along a mid-level field line in M06.  


The paper is organized  as follows.
In \S 2, we outline the GRMHD formulation and the wind equation (WE). 
In \S 3, with the  consideration of the conservation of energy flux in inflow and outflow region near the separation surface, we discuss the matching condition to connect the inflow and outflow part of a  PFD GRMHD flow.
In \S4, we introduce our model setup. We adopt similar parameters to those reported by M06 and compare the solution obtained by the matching condition with that of the time-averaged GRMHD numerical simulation results in M06.
Finally, a summary is given in \S5.

\section{Stationary, axisymmetric MHD Flow in Kerr Spacetime}
\label{sec:MHD} 

\subsection{Basic Formulae} 
The theory about stationary and axisymmetric ideal GRMHD flows has been in several works \citep{C86a,C86b,C87,F96,F04,T90,FG01}. For completeness, in this section we summary and present the necessary formulae for the purpose of this paper.

The natural unit system is used throughout this work. 
As $c=G=M=1$,  the gravitational radius $r_{\rm g} = GM/c^2  = 1$, 
   where $c$ is the speed of light, $G$ is the gravitational constant, and $M$ is the mass of the black hole 
(conversions from the c.g.s.\ units to the natural units for the physical variables here can be found 
  in tables 3 and 4 in \citet[][]{pu12}). 
The flows occur in a background Kerr space-time, 
  which is stationary and axisymmetric.  
For a metric signature $[-~+~+~+]$, the Kerr metric (in Boyer-Lindquist coordinates) reads 
\begin{eqnarray} 
ds^{2} & = & -\frac{\Delta-a^{2}\sin^{2}\theta}{\Sigma}~dt^{2}-\frac{4ar\sin^{2}\theta}{\Sigma}~dtd\phi  \nonumber \\ 
  & &  +\frac{A\sin^{2}\theta}{\Sigma}~d\phi^{2}+\frac{\Sigma}{\Delta}~dr^{2}+\Sigma~d\theta^{2}\; ,
\end{eqnarray}
where $a\equiv J$ is the angular momentum of the black hole, 
  $\Delta\equiv r^{2}-2r+a^{2}$, 
  $\Sigma\equiv r^{2}+a^{2}\cos^{2}\theta$, 
  and $A\equiv(r^{2}-a^{2})^{2}+\bigtriangleup a^{2}\sin^{2}\theta$. 

We also assume that the flow is cold. 
For a highly-relativistic flow, 
  the thermal pressure $p$ is insignificant 
  compared with the rest-mass energy density and the kinetic energy density in the fluid, 
  and hence the cold limit is justified.

The flow is magnetized and the stress-energy tensor of the fluid has two components:  
\begin{equation} 
   T^{\mu\nu}=T^{\mu\nu}_{\rm FL}+T^{\mu\nu}_{\rm EM} \ ,  
\end{equation} 
   where the fluid component is given  by 
\begin{equation}
  T^{\mu\nu}_{\rm FL} = \rho u^\mu u^\nu \  ,
\end{equation}
  and the electromagnetic component by 
\begin{equation}
  T^{\mu\nu}_{\rm EM} = \frac{1}{4\pi}
                \left(F^{\mu\gamma}F^{\nu}_{\gamma}-\frac{1}{4}g^{\mu\nu}F^{\alpha\beta}F_{\alpha\beta}\right)\; ,  
\end{equation}
   where $u^{\mu}$ is the 4-velocity of the fluid and 
  $\rho$ is the rest-mass energy density. 
The electromagnetic field tensor $F_{\mu\nu}$ satisfies Maxwell's equations, 
  and the proper number density $n$ 
  ($=\rho/m$, where $m$ is the rest-mass of the particles)  
  satisfies the mass continuity equation.


Under the ideal MHD condition, a stationary and axisymmetric flow obeys four conservation laws:   
\begin{eqnarray} 
 \label{eq:con1} 
   (n u^{\mu})_{;\mu}  & = & 0   \  ,  \\  
\label{eq:con2}
   F_{\mu\nu}u^{\mu} & = & 0   \  .    \\
 \label{eq:con3}
   (\xi_{\mu}T^{\mu\nu})_{;\nu} & = & 0 \  ,  \\   
 \label{eq:con4}
   (\eta_{\mu}T^{\mu\nu})_{;\nu}& = & 0 \  ,
\end{eqnarray} 
  where $\xi^{\mu}=\partial/\partial t$ 
  and $\eta^{\mu}=\partial/\partial \phi$ are the Killing vectors.   
These conservation equations 
  (equations \ref{eq:con1}--\ref{eq:con4})  
give four conserved quantities along a streamline.
By denoting the poloidal stream function as $\Psi (r,\theta)$, they are:
  (i) the angular velocity of the field line, $\Omega_{\mathrm{F}}(\Psi)$, 
  (ii) the particle number flux per unit magnetic flux  (mass loading), $\eta(\Psi)$, 
  (iii) the total energy of the flow per particle, $E(\Psi)$, and 
  (iv) the total angular momentum per particle, $L(\Psi)$ \citep[][]{C86a,C86b,C87,T90}:

\begin{eqnarray}
    \Omega_{\mathrm{F}} (\Psi) & = & \frac{F_{tr}}{F_{r\phi}}=\frac{F_{t\theta}}{F_{\theta\phi}}\   ,  
\end{eqnarray} 
\begin{eqnarray} 
\label{eq:eta_def}
    \eta (\Psi) & = & \frac{\sqrt{-g}\ nu^{r}}{F_{\theta\phi}} \ = - \ \frac{\sqrt{-g}\ nu^{\theta}}{F_{r \phi}}  
      \nonumber \\ 
     &  = & \frac{\sqrt{-g}\ nu^{t}(\Omega-\Omega_{\mathrm{F}})}{F_{r\theta}}   \  ,  
\end{eqnarray}
\begin{eqnarray}
\label{eq:E_def}
   E (\Psi) &=&E_{\rm FL}+E_{\rm EM} \nonumber \\
&=&-\mu u_{t}-\frac{\Omega_{\mathrm{F}}}{4\pi\eta}\sqrt{-g}\ F^{r\theta}\nonumber\\
&=&-\mu u_{t}-\frac{\Omega_{\mathrm{F}}}{4\pi\eta}B_{\phi}\  ,  
\end{eqnarray}
\begin{eqnarray}
\label{eq:L_def}
L (\Psi) &=&L_{\rm FL}+L_{\rm EM} \nonumber \\
&=&\mu u_{\phi}-\frac{1}{4\pi\eta}\sqrt{-g}\ F^{r\theta}\nonumber\\
&=&\mu u_{\phi}-\frac{B_{\phi}}{4\pi\eta}  \   , 
\end{eqnarray}
  with $\sqrt{-g}=\Sigma\sin\theta$.  
Here, $\Omega = u^{\phi}/u^{t}$ is the fluid angular velocity and $\mu$ is the relativistic specific enthalpy, which becomes $m$ in the cold limit. 
The covariant magnetic field observed by a distant observer with $u^{\nu}=(1,0,0,0)$ is given by 
\begin{equation}\label{b_def}
  B_{\mu} \equiv \frac{1}{2}\epsilon_{\nu\mu\alpha\beta}F^{\alpha\beta}u^{\nu}\; ,
\end{equation} 
and its toroidal component is given by
\begin{equation} 
  B_{\phi} = \sqrt{-g}\ F^{r\theta}=(\Delta/\Sigma)\sin\theta F_{r\theta} \ ,   
\end{equation} 
where $\epsilon_{\nu\mu\alpha\beta} =\sqrt{-g}~[\nu\mu\alpha\beta]$ is the Levi-Civita tensor, 
and $[\nu\mu\alpha\beta]$ is the completely antisymmetric symbol (see Appendix).   
  
The outward energy flux in the flow is  
\begin{equation} 
  \mathcal{E}^{r} = -T_{t}^{r}=nEu^{r} \  , 
\end{equation}  
   and the outward angular momentum flux is 
\begin{equation} 
  \mathcal{L}^{r} = -T_{\phi}^{r}=nLu^{r} \ . 
\end{equation}   
Splitting them into fluid (i.e.\ $E_{\rm FL}$ and $L_{\rm FL}$) 
   and the electromagnetic components (i.e.\ $E_{\rm EM}$ and $L_{\rm EM}$) gives  
\begin{eqnarray}
\label{eq:Eflux_def}
    \mathcal{E}^{r} & = & \mathcal{E_{\rm FL}}^{r}+\mathcal{E_{\rm EM}}^{r}   \nonumber   \\
                            & =  & nE_{\rm FL}u^{r}+nE_{\rm EM}u^{r}  \nonumber   \\
                            &=&-n\mu u_{t}u^{r}-\frac{\Omega_{\rm F}}{4\pi}\frac{B_{\phi}}{\Sigma\sin\theta}A_{\phi,\theta}\;,
\end{eqnarray} 
  and 
\begin{eqnarray}
\label{eq:Lflux_def}
   \mathcal{L}^{r} & = & \mathcal{L_{\rm FL}}^{r}+\mathcal{L_{\rm EM}}^{r}   \nonumber  \\
                            & = & nL_{\rm FL}u^{r}+nL_{\rm EM}u^{r} \nonumber  \\
                            &=&-n\mu u_{\phi}u^{r}+\frac{\mathcal{E}_{\rm EM}^{r}}{\Omega_{\rm F}}\;.
\end{eqnarray} 
As initially proposed by \citet{T90}, the case of $\mathcal{E}^{r}>0$ for inflow (which requires a negative total energy) is known as the MHD Penrose
process. For later studies, the term is instead used to 
indicate a negative energy orbit of the fluid component, $\mathcal{E}_{\mathrm{FL}}^{r}>0$
\citep[e.g.][]{H92,K02,S04,K05,K14}.

The bulk Lorentz factor of the flow for a distinct observer can be defined by 
\begin{equation}\label{eq:Gamma_def}
   \Gamma=\sqrt{-g_{tt}}\ u^{t}\  .  
\end{equation}
If all the energy in the Ponyting flux is converted to the fluid's bulk (kinetic) energy at a large distance,
the terminal Lorentz factor will be   
\begin{equation}\label{eq:lorentz_max}
\Gamma_{\infty}=\frac{E}{\mu}=\frac{\mathcal{E}^{r}}{\rho u^{r}}\ . 
\end{equation}
In addition, the angular velocity of the fluid at a large distance will be  
\begin{equation}\label{eq:angularv_max}
   u^{\phi}_{\infty} = \frac{L}{\mu}=\frac{\mathcal{L}^{r}}{\rho u^{r}} \  .
\end{equation}
Equations (\ref{eq:lorentz_max}) and (\ref{eq:angularv_max}) therefore provide the upper limit of the terminal Lorentz factor and angular velocity of the fluid at large distances.

\subsection{Wind Equation} 
The streamline of the flow is represented by the function $\Psi (r,\theta) ={\rm const}$.    
The WE (i.e.,  the relativistic Bernoulli equation),  
  describing the fluid motion along the streamlines   
  can be obtained using the normalization condition $u^{\alpha}u_{\alpha}=-1$. 
The WE therefore has the form:    
\begin{equation}
\label{eq:WE1}
   u_{\rm p}^{2}+1 = \left(\frac{E}{\mu}\right)^{2}\ U_{g}(r,\theta)  \    ,  
\end{equation} 
  where the poloidal component of the 4-velocity is given by 
 \begin{equation}
 \label{eq:up_def}
   u_{\mathrm{p}}^{2} =  u^{j}u_{j}  \ ,  
\end{equation}  
  with the summation over the poloidal indices  $j = \{r, \ \theta\}$.  
The term $U_{g}(r,\theta)$ in the right-hand side of equation (\ref{eq:WE1}), which is evaluated along the magnetic field line in the calculation,  
  is related to the conserved quantities, and its explicit expression depends on the assumed background space-time 
  (see \citet{C86a,C86b,C87,F96,F04} for the Minkowski and Schwarzschild spacetimes; and  
   \citet{T90,FG01} for the Kerr space-time).  
   
In a Kerr space-time, we obtain
\begin{equation}\label{eq:ug}
  U_{g}(r,\theta)=\frac{K_{0}K_{2}-2K_{2}M_{\rm A}^{2}-K_{4}M_{\rm A}^{4}}{(M_{\rm A}^{2}-K_{0})^{2}} \   , 
\end{equation}  
  \citep{T90}, where 
\begin{eqnarray}
   K_{0} & = &  -(g_{\phi\phi}\Omega_{\mathrm{F}}^{2}+2g_{t\phi}\Omega_{\mathrm{F}}+g_{tt})  \ ,  \\ 
    K_{2}  & = &   \left(1-\Omega_{\mathrm{F}}\frac{L}{E}\right)^{2}  \ ,   \\ 
    K_{4}  & = &  \frac{-1}{g_{t\phi}^{2}-g_{tt}g_{\phi\phi}}  
       \left[g_{\phi\phi}+2g_{t\phi}\frac{L}{E}+g_{tt}\left(\frac{L}{E}\right)^{2}\right]  \ .  
\end{eqnarray} 
The  Alfv\'en Mach number $M_{\rm A}$ is given by 
\begin{equation}
\label{eq:mach_def2}
   M_{\rm A}^{2}\equiv 4\pi\mu\eta^{2}/n = 
    4\pi\mu n\left(\frac{u_{\mathrm{p}}}{B_{\mathrm{p}}}\right)^{2} = 
    4\pi\mu\eta\left(\frac{u_{\mathrm{p}}}{B_{\mathrm{p}}}\right)  \  ,   
\end{equation} 
  where the re-scaled poloidal field  is
\begin{eqnarray}
\label{eq:bp} 
 B_{\mathrm{p}}^{2} & = &  B^{j}B_{j}\left( g_{tt}+g_{t\phi}\Omega_{\mathrm{F}}\right)^{-2}    \nonumber \\ 
 & = &  \frac{1}{\Delta\sin^{2}\theta}\left[~g^{rr}\Psi_{,r}^{2}+g^{\theta\theta}\Psi_{,\theta}^{2}\right]  \   .  
 \end{eqnarray} 
 
 Along the streamline several characteristic surfaces can be defined. Their definition and properties are summarized in the Appendix.
 
The conserved quantities $E$, $L$ and $\eta$ 
  can be expressed in terms of three system parameters: 
  (i) the launching point of the flow, $r_{\star}$,    
  (ii) the location of the Alfv\'en surface, $r_{\rm A}$, and 
  (iii)  the magnetization parameter at the launching point, $\sigma_{\star}$. 
Explicitly, the relations are  
\begin{eqnarray}
   \frac{L}{E} & = & -\frac{g_{t\phi}+g_{\phi\phi}\Omega_{F}}{g_{tt}+g_{t\phi}\Omega_{F}}\bigg\vert_{r_{\rm A}}  \ , 
\label{eq:LdE}  \\ 
   \tilde{E}^{2} & = & \frac{K_{0}}{K_{2}}\bigg\vert_{r_{\star}}   \ ,   
\label{eq:E_inj}  \\ 
\sigma_{\star} &  = &   \frac{\Phi_{\star}^{2}}{4\pi m(n u_{\rm p}\sqrt{-g})|_{r_{\star}}} 
  =\frac{\Phi_{\star}}{4\pi m}\frac{1}{|\eta|}   \  ,   
\label{eq:sigma_def}
\end{eqnarray}
where $\tilde{E}=E/\mu$, the flux function $\Phi = B_{\rm p}\sqrt{-g}$, and $\Phi_{\star}$ denotes that it is evaluated at $r=r_{\star}$. 
In terms of these parameters, the Mach number can be written as
\begin{equation}
   M_{\rm A} = \left[ \frac{\mu}{m}u_{\rm p}\sqrt{-g}~\frac{f}{\sigma_{\star}} \right]|^{1/2}    \  ,
\end{equation}
   where $f= \Phi_{\star}/\Phi$. 
Note that 
$u_{\rm p}=0$ at $r=r_{\star}$ has been assumed in deriving the relation (\ref{eq:E_inj}).  
In addition, the relation (\ref{eq:sigma_def}) implies that  knowing the mass loading $\eta$
   is equivalent to knowing the  $\sigma_{\star}$.

In the cold limit, 
  the WE is a polynomial equation of 4-th order in $u_{\rm p}$:  
\begin{equation}\label{eq:coldWE}
   \sum_{i=0}^{4} {\cal A}_{i} u_{\rm p}^{i}=0 \ . 
\end{equation}
The coefficients ${\cal A}_{i}(r;\Psi,\Phi,r_{\star},r_{\rm A},\sigma_{\star},\Omega_{\rm F})$ are given by 
\begin{eqnarray}
 {\cal A}_{4}  & = &   C_{2}^{2}   \ ,   \nonumber \\
 {\cal A}_{3} & = &   -2C_{2}K_{0}  \ ,    \nonumber \\
 {\cal A}_{2} & = &   C_{2}^{2}+K_{0}^{2} +\tilde{E}^{2}K_{4}C_{2}^{2}  \nonumber \\ 
 {\cal A}_{1} & = &   -2C_{2}K_{0} + 2\tilde{E}^{2}K_{2}C_{2}  \nonumber \\ 
 {\cal A}_{0} & = &   K_{0}^{2} - \tilde{E}^{2}K_{0}K_{2} \nonumber  \ , 
\end{eqnarray}
 where $C_{2}=\sqrt{-g}\ f/\sigma_{\star}$  \citep[][]{FG01}.


\section{Matching Condition of The Inflow and Outflow}
\label{sec:matching} 
In the work of \citet[][]{Phi83}, the matching of the inflow and outflow parts of the flow is constrained by the conservation of the  energy flux per magnetic flux in the inflow and outflow region 
\begin{equation}\label{eq:etae}
(\eta E)_{\rm inflow} =(\eta E)_{\rm outflow}
\end{equation}
 Remember that both $\eta$ and $E$ of the inflow and outflow are constant.

Consider equation (\ref{eq:etae}) at the separation surface, $r_{\rm s}$ ,  for PFD flow ($E\approx E_{\rm EM}\gg E_{\rm FL}$), we further consider 
\begin{equation}\label{eq:etaem}
(\eta)_{\rm inflow} (E_{\rm EM})^{-} = (\eta)_{\rm outflow} (E_{\rm EM})^{+}\;,
\end{equation}
to be the matching condition of the inflow and outflow.
The superscripts ``$-$"  (or ``$+$"), respectively, denote the physical value computed at the location very close to $r_{\rm s}$ in the inflow (or outflow) region, that is, 
 $r\to r_{\rm s}^{-}$ (or $r\to r_{\rm s}^{+}$).
After some algebra, equation (\ref{eq:etaem})
can also be expressed as
\begin{equation}\label{eq:ers}
(\mathcal{E}^{r}_{\mathrm{EM}})^{-} =(\mathcal{E}^{r}_{\mathrm{EM}})^{+}\;,
\end{equation}
or   
\begin{equation}\label{eq:bs}
\frac{\Omega_{\rm F}}{4\pi}(B_{\phi})^{-} = \frac{\Omega_{\rm F}}{4\pi}(B_{\phi})^{+}\;.
\end{equation}
Equation (\ref{eq:ers}) implies that the matching condition we adopt is equivalent to the statement: the outward Poyting energy flux is continuous at the separation surface\footnote{{\em cf.}
equation (\ref{eq:etae}) gives $(\mathcal{E}^{r})^{-} =(\mathcal{E}^{r})^{+}\;$.}. 
Equation (\ref{eq:bs}) revels that such condition guarantees that the toroidal field is continuous at the separation point, provided that $\Omega_{\rm F}$ is the same constant in the inflow and outflow region.

It is interesting to note that the matching condition does not require that $\eta$ or $E_{\rm EM}$  should be continuous when crossing $r=r_{\rm s}$.
That is, if we define 
\begin{equation}\label{eq:delta_def}
\delta\equiv\left|\,\,\frac{(\eta)_{\rm inflow}}{(\eta)_{\rm outflow}}\,\,\right|=\left|\,\,\frac{(E_{\mathrm{EM}})^{+}}{(E_{\mathrm{EM}})^{-}}\,\,\right|
=\left|\,\,\frac{(\sigma_{\star})^{+}}{(\sigma_{\star})^{-}}\,\,\right|\;,
\end{equation}
$\delta$ is not necessary for unity. The last relation in equation (\ref{eq:delta_def}) is obtained with the help of equation (\ref{eq:sigma_def}). 
Nevertheless, due to following reason, the outflow can still be properly constrained by the inflow even with the uncertainty of $\delta$.

Consider a flow along a prescribed, hole-threading poloidal field line with some specific angular velocity field.
\citet[][]{zna77} showed that, due to the regularity requirement at the event horizon, $r_{+}=1+\sqrt{1-a^{2}}$,
the derivative of the stream function 
$\Psi$ is finite and $B_{\phi}$
satisfies
\begin{equation}
B_{\phi}(r_{+},\theta)=-\sin\theta\frac{r_{+}^{2}+a^{2}}{r_{+}^{2}+a^{2}\cos^{2}\theta}(\Omega_{H}-\Omega_{F})\Psi_{,\theta}\;,
\end{equation}
where $\Omega_{H}$ is the angular velocity of the hole.
As a result, $(B_{\phi})^{-}$ is insensitive to different value of  $(\sigma_{\star})^{-}$ ($(B_{\phi})^{-}\approx const.$).
From dynamical point of view, this can be understood by the fact that the fast point of a PFD GRMHD inflow is always located close to the black hole event horizon (Appendix).

For outflow, however, there is no constraint at infinity, and therefore $(B_{\phi})^{+}$ depends on $(\sigma
_{\star})^{+}$ more strongly. 
Again, from dynamical point of view, the relatively strong dependence can be understood by the fact that the fast point of the outflow can vary from finite distance to infinity. 
Because the uncertainty of the $\delta$ is introduced by the uncertainty of $(\sigma_{\star})^{-}$, instead of $(\sigma_{\star})^{+}$ (see also \S4.2), the outflow can still be well constrained. 
The matching condition then play the role to constrains the outflow by singling out the outflow solution that satisfies $(B_{\phi})^{-}=(B_{\phi})^{+}$.

\section{Flow along a parabolic field line with a finite-distance fast point}
\label{sec:setup}
\subsection{Model Setup}

In general, the field configuration should be consistently determined by solving the trans-field equation, (i.e. the Grad-Shafranov equation). The trans-field  equation in cold limit involves the stream function $\Psi$, and  the derivative of the conserves quantities, $d\Omega_{\rm F}/d\Psi$, $d\eta/d\Psi$, $dE/d\psi$, $dL/d\psi$ \cite[][]{Ni91,Be93}.
However, solving the trans-field equation analytically is very challenging and beyond the scope of this paper.

On the other hand, we are interested in the case where  the fast point of the outflow is located at a finite distance.
It is therefore essential to consider an additional modification on the original force-free field line due to the MHD flow.  
We leave a better consideration of field configuration for a future work, and adopt the streamline function in \citet[][]{BN06} as the prescribed parabolic field
\begin{equation}\label{eq:change2}
\Psi=\Psi_{0}+\epsilon f\;,
\end{equation}
where  $\Psi_{0}$ is the the flat spacetime parabolic force-free field generated by the toroidal surface current distribution, $I=C/4\pi r(1+\Omega_{\rm F}r^2)^{1/2}$, on equatorial plane 
\citep[][]{B76,Le04},
\begin{equation}\label{eq:ff_Lee}
\Psi_{0}=\frac{\pi C}{\Omega_{\rm F}}\sinh^{-1}(\Omega_{\rm F} r (1-cos\theta))\;,
\end{equation}
and 
\begin{equation}
\epsilon f=\epsilon\; \pi C\Omega_{\rm F}r\sin\theta\,\;,\,\,\,\,\,\epsilon\ll1
\end{equation}
is the perturbation introduced by the MHD effect. 
The constant $C$ is assumed to be unity. 
Note that  by the help of the relation $\sinh^{-1}(x)=\ln(x+\sqrt{x^2+1})$, $\Psi_{0}$ is proportional to $r(1-cos\theta)$, which is the same as the dominating term of the parabolic field\footnote{Due to a similar  toroidal surface current distribution, $I=C/4\pi r$, on the equatorial plane, the parabolic force-free field around a black hole considered in BZ77: 
$\Psi=\frac{C}{2}\left\{ r(1-\cos\theta)+2\left(1+\cos\theta\right)\left[1-\ln\left(1+\cos\theta\right)\right]\right\}$,
also follows $\Psi\propto r(1-cos\theta)$ at large distance. 
This is one of the solutions of the source-free Maxwell equation in Schwarzschild spacetime: $(\frac{(1-2/r)}{\sin\theta}\Psi_{,r})_{,r}+(\frac{1}{r^{2}\sin\theta}\Psi_{,\theta})_{,\theta} =0$; while Equation (\ref{eq:ff_Lee}) is one of the solutions of the source-free Maxwell equation in flat spacetime: $(\frac{1}{\sin\theta}\Psi_{,r})_{,r}+(\frac{1}{r^{2}\sin\theta}\Psi_{,\theta})_{,\theta} =0$.}
in BZ77.
In addition, $\nabla\cdot B=0$ is guaranteed. It is shown in \citet[][]{BN06} that, although $\epsilon f/\Psi_{0}\ll1$ on the (outer) fast surface, the perturbation method  is not applicable beyond the fast point. As a result, we can only discuss the flow solution up to the outer fast point.


\begin{figure*}
\begin{center}
\includegraphics[scale=0.7,trim=0.cm 0.cm 0cm 0cm, clip=true,width=0.8 \textwidth  ]{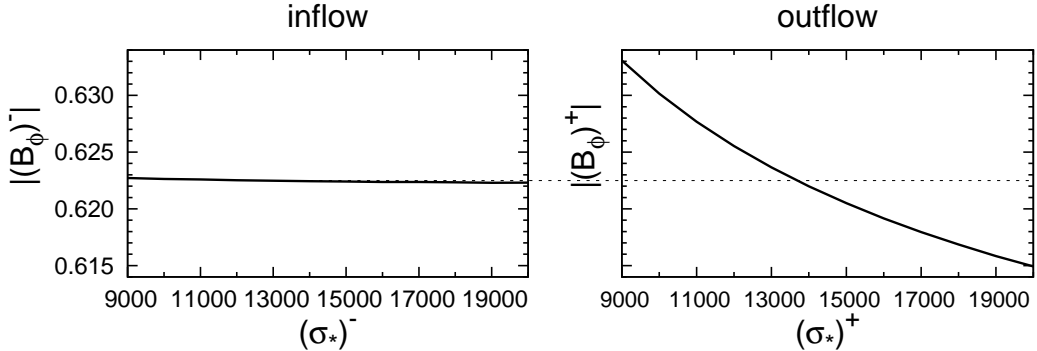}
\caption{The toroidal field, $B_{\phi}$,  as a function of magnetization parameter, $\sigma_{\star}$, near the separation surface, $r_{\rm s}$ for the inflow (left) and outflow (right) part of the solution. The superscript ''$-$" and ''$+$" denote the value computed at $r\to r_{\rm s}^{-}$ and $r\to r_{\rm s}^{+}$, respectively.
The matching condition constrain the outflow by the inflow; that is, singling out the outflow solution that satisfy $(B_{\phi})^{+}=(B_{\phi})^{-}$ (see \S3.2 and \S4.2). When $\delta=1$ ($(\sigma_{\star})^{-}=(\sigma_{\star})^{+}$$\simeq13700$), the matching condition is always satisfied.} \label{fig:matching}
\end{center}
\end{figure*}


The following assumptions for a PFD GRMHD flow along a hole-threading field line are  considered.
First, we assume $r_{\star}=r_{\rm s}$. The assumption of  $r_{\star}=r_{\rm s}$ and $u_{\rm p}|_{r_{\star}}=0$ ensures $u_{\rm p}$ has a smooth  transition from $u_{\rm p}<0$ (inflow) to $u_{\rm p}>0$ (outflow).
Second, to guarantee the flow is PFD, we require $\sigma_{\star}\gg 1$ (see also \S3.4.1 of \citet[][]{FG01} for an estimation) and $\epsilon\ll 1$.
Furthermore, we assume $\epsilon$ is constant along field lines. The higher the value of $\sigma_{\star}$, the more magnetically dominated the flows are.

Among all the parameter space we seek for the parameter set $\{\Omega_{\rm F},\eta,E,L\}$, which gives a similar time-averaged GRMHD simulation result in M06 for comparison.
We therefore focus on a spinning black hole with its dimensionless spin $a=0.9$ and a field line that threads the event horizon at mid-latitude, $\theta=60^{\circ}$. As mentioned in \S2.2, the set $\{\Omega_{\rm F},\eta,E,L\}$ can be equivalently determined by $\{\Omega_{\rm F},\sigma_{\star},r_{\star},r_{\rm A}\}$.  We adopt $r_{\star}=r_{\rm s}$ (assumed), and $\Omega_{\rm F}=1/2 \;\Omega_{\rm H}$ (similar to the result of M06) in both inflow and outflow region. Then we determine $\sigma_{\star}$ and $r_{\rm A}$ (note that once $r_{\rm A}$ is determined, the location of the fast surface is determined accordingly) by the constraint of: (i)
$(E_{\rm EM}/E_{\rm FL})\sim10^{2}$ near the separation surface\footnote{In top panel of  Figure 7 of M06\, $\Gamma^{({\rm EM})}_{\infty}/ \Gamma^{({\rm MA})}_{\infty}$ is $\sim 10^{3}$ (see M06 for definitions), which equivalent to $E_{\rm EM}/E_{\rm FL}\sim10^{2}$ in terms of the definition in this paper. Note that  our definition of $E_{\rm EM}$ is $4\pi$ smaller than $\Gamma^{(\rm EM)}_{\infty}$ used in M06. The factor of $4\pi$ is absorbed into the definition of $F^{\mu\nu}$ in M06.} similar to the case in M06, and (ii) the matching condition.

We note that $\Omega_{\rm F} \approx1/2 \;\Omega_{\rm H}$ 
is self-consistently
obtained in a steady PFD GRMHD flow solution for a monopole field geometry \cite[][]{BK00},
while it may not be relevant for a parabolic field geometry. BZ77 examined the parabolic streamline in which $\Omega_{\rm F}$
decreases when shifting the angle from close to the pole to equatorial
plan. 
In M06, the field geometry becomes almost monopolar in the
vicinity of the horizon, so that $\Omega_{\rm F}\approx1/2 \;\Omega_{\rm H}$ is observed along the
field line \cite[see also][]{B09}. In the present paper, although the parabolic field is
prescribed as a global field geometry, we nevertheless adopt the
constant value of $1/2 \;\Omega_{\rm H}$ in our fiducial solution for convenience.

\subsection{Matching the Inflow and Outflow}

Consistent solutions for PFD inflow and outflow along a field line are obtained
iteratively until the matching condition is satisfied. 
For each $\sigma_{\star}$, because $\Omega_{\rm F}$ ($=\Omega_{\rm H}/2$) and $r_{\star}$ (determined by where $K'_0=0$ is along the field line; see Appendix) are known, we can solve WE (equation (\ref{eq:coldWE})) for  the flow solution by requiring the physical solution to pass through the fast surface. 
For example, for the case $\epsilon=0.065$, the profiles of $(B_{\phi})^{-}$ and $(B_{\phi})^{+}$ as a function of $(\sigma_{\star})^{-}$ and  $(\sigma_{\star})^{+}$, respectively, are shown in Figure \ref{fig:matching}.

A consistent inflow/outflow solution exists when a suitable set $(\epsilon,(\sigma_{\star})^{-},(\sigma_{\star})^{+})$ is applied.
As mentioned in \S3,   the tendency is for $(B_{\phi})^{-}\approx const.$ to result in multiple choices of $(\sigma)^{-}$ such that $(B_{\phi})^{-}=(B_{\phi})^{+}$ is satisfied. 
This leads to certain amount of freedom for choosing the value for $\delta$.
For simplicity, $\delta=1$ is assumed, so $(\sigma_{\star})^{-}=\sigma_{\star}|^{+}=\sigma_{\star}$. 
We can read from Figure \ref{fig:matching} that  $(B_{\phi})^{-}=(B_{\phi})^{+}$  is satisfied when $(\epsilon,\sigma_{\star})=(0.065, \simeq13700)$. 

By the same method, for any specific value of $\epsilon$ (or $\sigma_{\star}$), there is a corresponding $\sigma_{\star}$ (or  $\epsilon$) that satisfies the matching condition. The quantitative relation shows that, as $\sigma_{\star}$ increases, $\epsilon$ also increases. This implies that,  because there is more mass loading onto the field, the field progressively bunches up toward to the rotational axis of the black hole. 
Finally, after $\sigma_{\star}$ is chosen by the matching condition, the parameter set  $\{\Omega_{\rm F},\sigma_{\star}, r_{\star},r_{A}\}$ of the inflow/outflow part of the solution is uniquely determined.
The relaxation of the assumption $\delta=1$ is discussed at the end of \S4.3.2.


\subsection{Self-Consistent Inflow/Outflow Solution}
\label{sec:inflow_zoom} 
\subsubsection{Flow Properties}
We adopt the above parameter set $(\epsilon,\sigma_{\star})=(0.065, \simeq13700)$ as the fiducial model parameters, because the resulting flow solution satisfies our requirement $(E_{\rm EM}/E_{\rm FL})|_{r_{\star}}\sim10^{2}$ (\S4.1).
The conserved quantities, $\{\Omega_{\rm F},\eta,E,L\}$, of our fiducial flow solution are shown in Table 1.
The mass loading $\eta$ changes sign according to $u^{r}$ and $\Psi,_{\theta}$ (equation (\ref{eq:eta_def})) in inflow and outflow regions. Because the sign has no specific meaning, the absolute value $|\eta|$ is shown. By the assumption $\delta=1$, $|(\eta)_{\rm inflow}|=|(\eta)_{\rm outflow}|$ (equation (\ref{eq:delta_def})).

In the inflow region  ($u^{r}<0$), both $E<0$ and $L<0$ 
indicates that the energy and angular momentum of the black hole is extracted outward ($\mathcal{E}^{r}>0$  and $\mathcal{L}^{r}>0$).
 $E/\mu$ of the outflow gives the maximum possible value of the terminal Lorentz factor (equation (\ref{eq:lorentz_max})).
Although $(E_{\rm EM})^{-}=(E_{\rm EM})^{+}$  under the assumption $\delta=1$, the absolution value of $E=E_{\rm EM}+E_{\rm FL}$ for the inflow is slightly smaller than the value for the outflow. This is because,
in the inflow region, the fluid component $E_{\rm FL}$ (or $L_{\rm FL}$) has an opposite sign with the electromagnetic component $E_{\rm EM}$ (or $L_{\rm EM}$), partly canceling the electromagnetically extracted energy (or angular momentum); whereas in the outflow region, the fluid and the electromagnetic components have the same sign, both carrying the energy and angular momentum outward.
The general properties of different physical components for PFD inflows and outflows are provided in Table \ref{tab:flow properties}.


\begin{table}
\caption{Conserved Quantities of the Fiducial Flow Solutions}
\label{tab:flow properties}
\begin{centering}
\begin{tabular}{ccc} 
\hline
\hline
\multicolumn{3}{c}{a=0.9}\\
\multicolumn{3}{c}{($\epsilon=0.065$, $\sigma_{\star}\simeq13700$)$^{\dagger}$ with $\delta=1$}\\
\hline
 & Inflow  & Outflow \\
\hline
$|\eta|(\Psi)m$	&\multicolumn{2}{c}{$\simeq7\times 10^{-5}$}\\
$E(\Psi)/\mu$		& $\simeq-112$&$\simeq114$ \\
$L(\Psi)/\mu$		& $\simeq-720$&$\simeq724$ \\
$\Omega_{\rm F}(\Psi)$				&\multicolumn{2}{c} {$\simeq0.157$}\\

\hline
\end{tabular} \\
\end{centering}
Notes.\\
$^{\dagger}$  a consistent inflow/outflow solution is obtained when a suitable set $(\epsilon,\sigma_{\star})$ is applied, such that the matching condition is satisfied  (see \S3 and \S4.2).
\end{table}



\begin{table}
\caption{Properties of PFD GRMHD Flow Along the Same Hole-Threading Field Line}
\label{tab:flow properties}
\begin{centering}
\begin{tabular}{ccc} 
\hline
\hline
 & Inflow Solution & Outflow Solution \\
\hline
$u^{r}$				& $<0$&$>0$ \\
$u^{\theta}$		& $>0$&$<0$ \\
$u^{\phi}$			& $>0$&$>0$ \\
$u^{t}$				& $>0$&$>0$ \\
\vspace*{0.03cm} \\ 
$E=E_{\mathrm{FL}}+E_{\mathrm{EM}}$									& $<0$ &$>0$ \\
$E_{\mathrm{FL}}$				& $>0$ &$>0$ \\
$E_{\mathrm{EM}}$					& $<0$ &$>0$ \\
\vspace*{0.03cm} \\ 
$L=L_{\mathrm{FL}}+L_{\mathrm{EM}}$									& $<0$ &$>0$ \\
$L_{\mathrm{FL}}$				& $>0$ &$>0$ \\
$L_{\mathrm{EM}}$					& $<0$ &$>0$ \\
\vspace*{0.03cm} \\ 
$\mathcal{E}^{r}=\mathcal{E}_{\mathrm{FL}}^{r}+\mathcal{E}_{\mathrm{EM}}^{r}$ & $>0$ &$>0$ \\
$^{\dagger}\; \mathcal{E}^{r}_{\mathrm{FL}}$	& $<0$&$>0$ \\
$\mathcal{E}^{r}_{\mathrm{EM}}$	& $>0$ &$>0$ 
\vspace*{0.05cm} \\ 
\hline
\end{tabular} \\
\end{centering}
Notes.\\ $^{\dagger}$  For a stationary GRMHD inflow solution along a hole-threading field line, $\mathcal{E}^{r}_{\mathrm{FL}}<0$ is satisfied. In contrast, for an inflow along a non hole-threading field line during transient phase, $\mathcal{E}^{r}_{\mathrm{FL}}>0$ is possible \citep[e.g.][]{K02,K05}.\\
\end{table}

\begin{figure}
\begin{center}
\includegraphics[scale=0.4, trim=0cm 0cm 0cm 0cm, clip=true, width=0.45 \textwidth ]{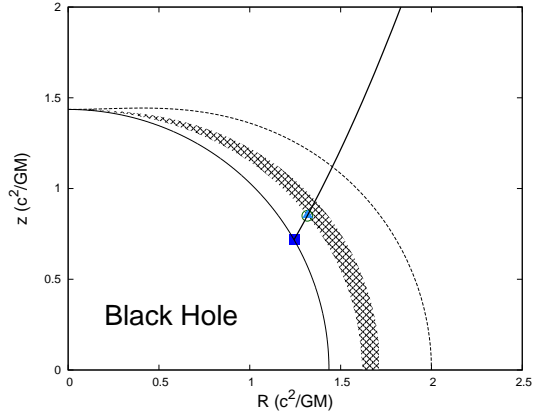}
\caption{Characteristic points of a fiducial PFD GRMHD inflow. Toward the black hole: Alfv\'en surfaces (filled cyan triangles), light surfaces (empty green circles), and fast surfaces (filled blue squares).
The event horizon and static limit (the outer boundary of the ergosphere) are shown by the thin solid and  dashed lines, respectively . 
Field lines are represented by the thick solid line.
The Alfv\'en surfaces are located inside the the negative energy region (shaded region), implying that the black hole energy is extracted outward.}
\label{fig:jet_line_zoom}
\end{center}
\end{figure}

The extraction of black hole rotation energy by the GRMHD inflow is also indicated by the location of the inflow Alfv\'en surface.
A remarkable feature in GRMHD is the existence of a negative energy region: once the Alfv\'en surface of an inflow resides inside such a region, the black hole energy is extracted outward \citep[][]{T90}.    
The inner boundary of the negative energy region is the inner light surface,  
  and the outer boundary is defined by $g_{tt}+g_{t\phi}\Omega_{\rm F}=0$.  
Thus, the region must be inside the ergosphere, where $g_{tt}>0$.  
As the flow becomes increasingly PFD, the location of the Alfv\'en surface moves toward  the light surface, finally entering the negative energy region (see the Appendix).
For PFD GRMHD inflow, the fast surface is located very close to, and almost coincides with, the black hole event horizon.  
This is why the PFD inflow solutions are all similar, as mentioned in \S3.
In Figure \ref{fig:jet_line_zoom} we plot the locations of the Alfv\'en surface and fast surface of the flow, which share the same features mentioned previously.


\subsubsection{Radial Structure}
\begin{figure*}
\begin{center}
\includegraphics[scale=1.,trim=0.3cm 0.2cm 0cm 0cm, clip=true,width=0.5\textwidth]{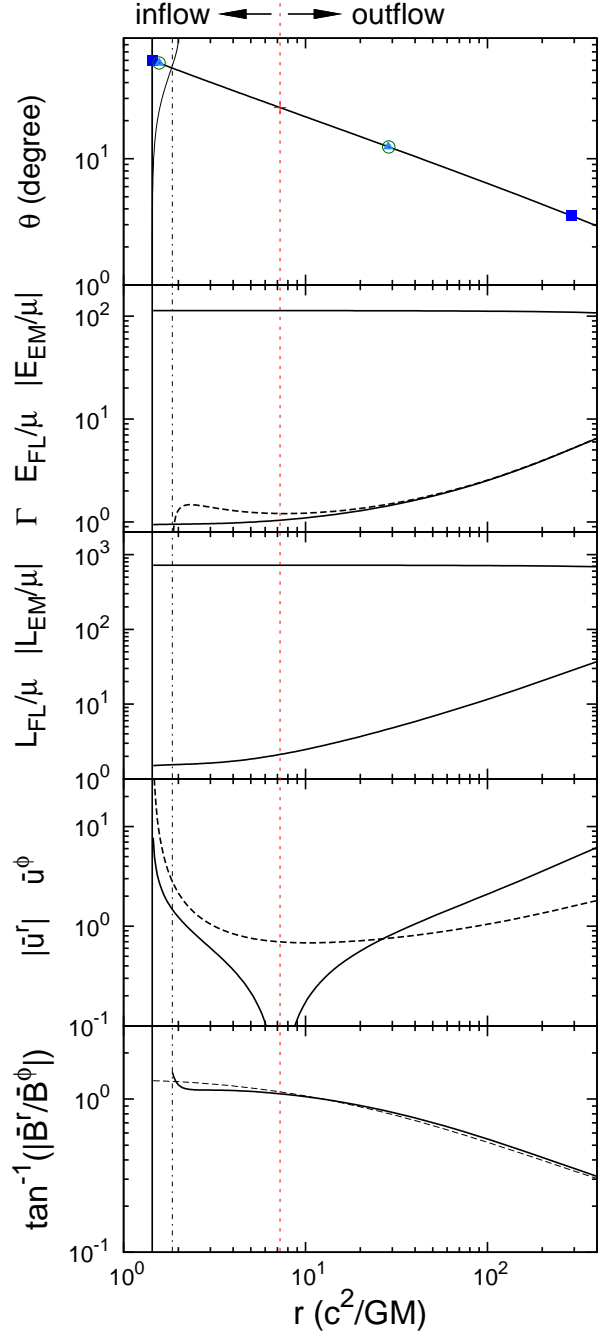}
\caption{Fiducial PFD GRMHD flow solution properties along a field line. {\em  Top panel:} Jet opening angle of the prescribed field. Location of characteristic surfaces are also shown: separation point (plus sign), light surfaces (empty circles), Alfv\'en surfaces (filled triangles), and fast surfaces (filled squares). The thin vertical line indicates the angular profile of the static limit ($g_{tt}=0$).
{\em  Second panel:} Electromagnetic energy component (upper solid line), $E_{\rm EM}$, and fluid energy component (lower solid line), $E_{\rm FL}$, of the total energy $E=const.=E_{\rm EM}+E_{\rm FL}$, in unit of fluid rest mass energy.
The profile of $\Gamma$ is shown only when $g_{tt}<0$ (dashed line).
{\em  Third panel:} Electromagnetic (upper solid line), $L_{\rm EM}$, and fluid (lower solid line), $L_{\rm FL}$ components of total angular momentum, $L=const.=L_{\rm EM}+L_{\rm FL}$. {\em  Fourth panel:} The orthonormal velocities $\bar{u}^{r}$ and $\bar{u}^{\phi}$. 
{\em Bottom panel:} The pitch angle of the orthonormal field, $\tan^{-1}(|\bar{B}^{r}/\bar{B}^{\phi}|)$ (solid line), which is well-described by $\tan^{-1}| R_{L}/\sqrt{g_{\phi\phi}}|$ (dashed line) at large distance, where $R_{\rm L}=1/\Omega_{F}$. Because the orthonomal field is related to $g_{tt}$ and becomes ill-defined near the black hole, the pitch angle is only shown when $g_{tt}<0$. Along the field line, the location of the event horizon,  the static limit, and the separation point are indicated by the vertical solid, dotted-dashed, and  dashed lines, respectively. }
\label{fig:single}
\end{center}
\end{figure*}

Let us now show the fiducial flow solution up to the fast surface in Figure \ref{fig:single} and compared the result (especially Figures 7 and 8) in M06.
The top panel of Figure \ref{fig:single} shows 
  the opening angle of the prescribed field, which roughly follows a single power law $\theta\propto r^{-0.52}$, which is in general more collimated compare to the result of M06.
  The locations of the characteristic surfaces are overlap onto the profile. The Alfv\'en surfaces are located close to the light surfaces, and the inner fast surface is located close to the horizon.
Note that in M06 the opening angle has different slope at a different radial range (see Figure 10 of M06). Instead, our prescribed field line follows a single power law. Nevertheless, with similar requirements at the separation surface ($(E_{\rm EM}/E_{\rm FL})|_{r_{\star}}\sim10^{2}$),
the fast surface of the outflow is located at several hundred $r_{g}$ from the black hole, which is similar to the result of M06.

The second panel of Figure \ref{fig:single}
  shows the profiles of the electromagnetic energy component $E_{\rm EM}$ 
  and the fluid energy component $E_{\rm FL}$ (both normalized by $\mu$), 
  and the Lorentz factor $\Gamma$. 
Inside the ergosphere, $g_{tt}<0$, $\Gamma$ is ill-defined.
Therefore, only the profile segments outside the ergosphere are plotted. 
At large distances, $g_{tt}\to -1$, $\Gamma\to -u_{t}=E_{\rm FL}/\mu$. 
In addition, for a PFD flow, $E\approx E_{\rm EM}\gg E_{\rm FL}$ when launching, so the maximum possible value of the terminal Lorentz factor, $\Gamma_{\infty}=E/\mu\approx E_{\rm EM}/\mu$ near the separation surface.
As a result, the profile of $\Gamma$ along the streamline is therefore related to the conversion from $E_{\rm EM}$ to $E_{\rm FL}$. 
In the acceleration region ($\gtrsim50r_{g}$), $\Gamma$ roughly follows $\propto r^{0.6}$, which is similar to the result of M06, and the analytical result of $\Gamma\propto r^{0.5}$ obtained in \citep[][]{BN06}. It is expected that a further acceleration is expected to be take place beyond the fast surface due to the magnetic nozzle effect \citep[e.g.][]{C89,LCB92}.
The conversion efficiency from Poynting to kinetic energy, which can be approximated by $\Gamma/\Gamma_{\infty}$, is closely related to the location of the fast surface.
For example, when the fast surface is located at infinity, $\Gamma/\Gamma_{\infty}\approx0$.
For the outflow solution, $\Gamma/\Gamma_{\infty}\lesssim0.1$ up to the fast surface, which is located at $\sim 300 r_{g}$.
It is also interesting to note that the flow has already reached modest Lorentz factors ($\Gamma\sim5$) at the fast surface, and most of the Poynting energy has not yet been converted to kinetic energy.
Note that, despite the final value of $\Gamma$ at the fast surface is similar to the result in M06, the  Poynting energy in M06 at fast surface has already experienced a significant decay (more than one order of magnitude) up to the fast surface. The reason why the fluid energy is not correspondingly increasing may be due to dissipative processes.
In the inner region beneath the separation surface,  
   $-u_{t}= E_{\rm FL}/\mu \lesssim 1$,
   as expected because the fluid is strongly bounded by the black hole's gravity.  
In the outer region beyond the separation surface, $-u_{\rm t} > 1$,  
   which implyies that the fluid is unbound and an outflow occurs.

Similar to the energy conversion between the fluid and the electromagnetic components, the increase of the fluid component of the angular momentum $L_{\rm FL}$ is at the expense of the electromagnetic component of the angular momentum $L_{\rm EM}$.
The profiles of $L_{\rm FL}$ and $L_{\rm EM}$ (normalized by $\mu$) are shown in  the third panel of Figure \ref{fig:single}. 
Again,
the profile of the fluid component $L_{\rm FL}/\mu=u_{\phi}$ is consistent with result of M06, but the decreases of the Poynting component in the simulation are much larger than our semi-analytical solution.

The radial and polar components of the four-velocity of the flow, $u^{r}$ and $u^{\theta}$,  
  can be calculated from equations (\ref{eq:eta_def}) and  (\ref{eq:up_def}), 
  with $u_{\rm p}$ determined by the WE. 
The other two components of the four-velocity, $u^{t}$ and $u^{\phi}$, 
   can be obtained by solving
\begin{equation}
    -\mu(u_{t}+u_{\phi}\Omega_{F})=E-\Omega_{F}L \ , 
\end{equation} 
 subject to the normalisation $u^{\alpha}u_{\alpha} = -1$.  
The velocity components $u^{r}$ and $u^{\theta}$ change signs 
across $r=r_{\rm s}$, 
while the velocity components $u^{\phi}$ and $u^{t}$ remain positive in both the inflow and outflow regions.  
The angular velocity of the fluid, $\Omega=u^{\phi}/u^{t}$, 
  which follows the black hole's rotation, 
  is however always positive along the magnetic field line. 
At the separation surface, $u^{r}=u^{\theta}=0$, and hence $\Omega=\Omega_{\rm F}$.

The radial and toroidal components of the orthonormal velocity at large distance are given by 
\begin{eqnarray}
   \bar{u}^{r}& = & \sqrt{g_{rr}}~u^{r}  \ ,  \\ 
   \bar{u}^{\phi}& = & \sqrt{g_{\phi\phi}}~u^{\phi} \,
\end{eqnarray} 
as shown in the fourth panel of Figure \ref{fig:single}. 
The profile of $\bar{u}^{r}$ is quite similar to the result in M06, but $\bar{u}^{\phi}$ has a relatively steeper profile compare to the simulation result. We suppose this is related to the field configuration beyond the fast point, where we are not able discuss in current prescribed field configuration.

The orthonormal components of the magnetic fields at large distance can be defined by
\begin{eqnarray}
   \bar{B}^{r}& = & \sqrt{g_{rr}}~B^{r}  \ ,   \\ 
       \bar{B}^{\phi} & =& \sqrt{g_{\phi\phi}}~B^{\phi}\;.
\end{eqnarray}
Note that $\bar{B}^{r}$ is given initially when solving the WE, and $\bar{{B}}_{\phi}$, which is not initially known, can be determined after solving the WE.
The bottom panel of Figure \ref{fig:single} shows the profile of the pitch angle, $\tan^{-1}(|\bar{B}^{r}/\bar{B}^{\phi}|)$.
Because $B^{r}$ and $B^{\phi}$ are both functions of $g_{tt}$ (see the Appendix), they quickly decrease and change sign 
when entering the ergosphere ($g_{tt}>0$). As a result, $|\bar{B}^{r}|$ and $|\bar{B}^{\phi}|$ are ill-defined close to the black hole, and we
only plot the profile  in the region where $g_{tt}<0$. The reason why the pitch angle profile in M06 does not have this problem should be related to the definition of the field. The explicit form of the magnetic field we adopt is provided in the Appendix. Nevertheless, at far region (e.g. the outflow region), spacetime becomes more flat and the differences of the definition are less important, our result agrees with the result of M06.
The locations where $|\bar{B}^{r}|=|\bar{B}^{\phi}|$ are close to the light surface.
At a large distance,
$|\bar{B}^{r}|$ is well-described by $|\bar{B}^{r}|\approx|\bar{B}^{\phi}| R_{L}/\sqrt{g_{\phi\phi}}$, where $R_{\rm L}=1/\Omega_{F}$, as also obtained in M06.

At the end of this section, we discuss how would the flow solution would change if we adopt a $\delta$, which also satisfies the matching condition, but does not equal to unity. Keep in mind that the outflow solution is well constrained by the matching condition, and the uncertainty of $\delta$ is due to the degeneracy of the inflow solutions (\S3). As a result, the outflow solution will remain the same if a different value of $\delta$ is adopted. For the PFD GRMHD flow, because the location of the Alfv\'en surface is always located near the inner light surface and the fast surface is always located close to the horizon, the flow dynamics will therefore be similar. That is, $u^{r}, u^{\theta}, u^{\phi}, u^{t}$, and therefore $E_{\rm FL}/\mu=-u_{t}$ and $L_{\rm FL}/\mu=u_{\phi}$ will remain almost unchanged. In addition, $B^{r}$ (prescribed) and $B^{\phi}$  (constrained by the Znajek's condition on horizon described in \S3) will also remain similar.
The electromagnetic component, $E_{\rm EM}$ and  $L_{\rm EM}$, due to the dependence of  $\propto 1/\eta$, follow $|(E_{\rm EM})_{\rm inflow}/(E_{\rm EM})_{\rm outflow}|=|(L_{\rm EM})_{\rm inflow}/(L_{\rm EM})_{\rm outflow}|=|(\eta)_{\rm outflow}/(\eta)_{\rm inflow}|=1/\delta$.


\section{Summary}

A semi-analytical scheme is presented to investigate the cold, PFD GRMHD flow solution along a Kerr black hole-threading field. The continuity of the outward Poynting energy flux across the separation surface is used as the matching condition to connect the inflow and outflow parts of a PFD GRMHD flow solution.
We consider the parabolic field line of \citet[][]{BN06},  and therefore the resulting flow passes through all the critical points at a finite distance.

With similar black hole spin, angular velocity of the field, and magnetization at the separation surface, we are able to obtain a specific parameter set $\{\Omega_{\rm F},\sigma_{\star},r_{\star},r_{\rm A}\}$ that  gives inflow and outflow solutions in agreement with the time-averaged flow properties along a mid-level field line reported in the GRMHD  simulation of M06.

In this current work, due to the limitation of the prescribed field configuration, we can only discuss the flow solution up to the outer fast surface. As a future work, a better consideration of the field configuration could help to explore of the flow acceleration beyond the fast surface, where the major jet acceleration takes place. 

Compared to the GRMHD and the general relativistic force-free electrodynamics (GRFFE) \citep[e.g.][]{mn07} numerical simulation approaches, the semi-analytical approach provides a complementary understanding of the relativistic jets, in the sense that the numerical dissipative process is absent, and that the fluid component is included. 
The stationary solution obtained by the scheme can also be provided as a reference of the time-averaged GRMHD jet behaviour in numerical simulations.

\acknowledgments 
We thank the anonymous referee for helpful suggestions, which significantly improved the paper.
We also thank C. Fendt for helpful information about GRMHD outflow solutions under cold limit, and Z. Younsi for proofreading. K.H. is supported by the Formosa Program between National Science Council  
in Taiwan and Consejo Superior de Investigaciones Cientificas
in Spain, administered through grant number 
NSC100-2923-M-007-001-MY3.
Y.M. is supported by the Ministry of Science and Technology of Taiwan
under the grant NSC 100-2112-M-007-022-MY3 and MOST103-2112-M-007-023-MY3, and by ERC Synergy Grant {}``BlackHoleCam:
Imaging the Event Horizon of Black Holes".

\appendix
\section{Notes on the magnetic field}
Here we present the explicit form of the magnetic field. The covariant magnetic field defined in Equation (\ref{b_def})
\begin{equation}
  B_{\mu} \equiv \frac{1}{2}\epsilon_{\nu\mu\alpha\beta}F^{\alpha\beta}\xi^{\nu}\; ,
\end{equation} 
 can be alternatively written as  
\begin{equation}
  B^{\mu} \equiv \frac{1}{2}\epsilon^{\nu\mu\alpha\beta}F_{\alpha\beta}\xi_{\nu}\;,
  \end{equation}
where $\epsilon_{\nu\mu\alpha\beta} =\sqrt{-g}~[\nu\mu\alpha\beta]$, and $\epsilon^{\nu\mu\alpha\beta} =-\frac{1}{\sqrt{-g}}~[\nu\mu\alpha\beta]$, with $\sqrt{-g}=\Sigma\sin\theta$. Since $\xi^{\nu}=(1,0,0,0)$ and $\xi_{\nu}=(g_{tt},0,0,g_{t\phi})$, we can quickly read from the above definitions that $B_{t}=0$ but $B^{t}\neq0$.
  
The components of the magnetic field are therefore given by
\begin{equation}
B_{r}=\sqrt{-g}F^{\theta\phi},
\end{equation}
\begin{equation}
B^{r}=-\frac{g_{tt}+\Omega_{\rm F} g_{t\phi}}{\sqrt{-g}}F_{\theta\phi}\;,
\end{equation}
\begin{equation}
B_{\theta}=-\sqrt{-g}F^{r\phi}\;,
\end{equation}
\begin{equation}
B^{\theta}=\frac{g_{tt}+\Omega_{\rm F} g_{t\phi}}{\sqrt{-g}}F_{r\phi}\;,
\end{equation}
and
\begin{equation}
B_{\phi}=\sqrt{-g}F^{r\theta}\;,
\end{equation}
\begin{equation}
B^{\phi}=-\frac{g_{tt}}{\sqrt{-g}}F_{r\theta}\;.
\end{equation}
With the relations
\begin{equation}
F^{\theta\phi}=-g^{\theta\theta}\frac{g_{tt}+\Omega_{\rm F}g_{t\phi}}{\Delta\sin^{2}\theta}\;F_{\theta\phi}\;,
\end{equation}
\begin{equation}
F^{r\phi}=-g^{rr}\frac{g_{tt}+\Omega_{\rm F}g_{t\phi}}{\Delta\sin^{2}\theta}\;F_{r\phi}\;,
\end{equation}
\begin{equation}
F^{r\theta}=g^{rr}g^{\theta\theta}\; F_{r\theta}\;,
\end{equation}
one can check $B^{r}=g^{rr}B_{r}$, $B^{\theta}=g^{\theta\theta}B_{\theta}$, and $B^{\phi}=g^{\phi\phi}B_{\phi}$. Note that, despite  $F_{\mu\nu}$ is finite  at all region, $B^{\mu}$ is ill-defined near a Kerr black hole because $g_{tt}$ changes sign when entering the ergosphere.  
 
 At large distance, the metric become Minkowski spacetime in spherical coordinates,
 \begin{equation}
 ds^{2}=-dt^{2}+dr^{2}+r^{2}d\theta^{2}+r^{2}\sin^{2}\theta d\phi^{2}\;,
 \end{equation}
 and $\sqrt{-g}=r^{2}\sin\theta$. In this limit, the orthonormal field has the form
 \begin{equation}
 \bar{B}^{r}\equiv\sqrt{g_{rr}}B^{r}=\frac{1}{r^{2}\sin\theta}F_{\theta\phi}\;,
 \end{equation}
 \begin{equation}
 \bar{B}^{\theta}\equiv\sqrt{g_{\theta\theta}}B^{\theta}=-\frac{1}{r\sin\theta}F_{r\phi}\;,
 \end{equation}
  \begin{equation}
 \bar{B}^{\phi}\equiv\sqrt{g_{\phi\phi}}B^{\phi}=\frac{1}{r}F_{r\theta}\;.
 \end{equation}

\section{Characteristic Surfaces}
In the following we outline the characteristic surfaces of cold GRMHD flow, including the light surfaces, the separation surface, and the Alfv\'en and fast surfaces.
\subsection{Light Surfaces}
\label{sec:surface}
The surfaces defined by $K_{0}=0$ are the light surfaces. 
There are two light surfaces in a black-hole magnetosphere, the outer and the inner light surfaces.  
In the regions outside the light surfaces (where $K_{0}<0$)
   the fluid streams radially 
   so as to avoid the toroidal velocity exceeding the speed of light. 
The outer light surface is formed 
  in the same manner as the light cylinder in a pulsar magnetosphere, 
  but it does not necessarily have a cylindrical shape in a Kerr space-time. 
  The inner light surface is formed due to strong gravity.
  Only when the black hole and the field line are not rotating does the inner light surface coincide with the black hole event horizon.
\subsection{Separation Surface}
In the cold limit
the fluid acceleration  
  along a field line, $u'_{p}$ (where prime denotes the derivative along the flow streamline),   
  changes direction at a certain point.    
The location, $r_{\rm s}$, at which the change occur forms a separation surface \citep[][]{T90, H92}. 
The fluid, starting with negligible velocity at $r_{\rm s}$, 
   is accelerated inward inside the separation surface, creating an inflow. 
It is however accelerated outward outside the surface and develops an outflow.  

\begin{figure}[h]
\begin{center}
\includegraphics[trim=6cm 6.5cm 6cm 9cm, clip=true,width=0.45 \textwidth]{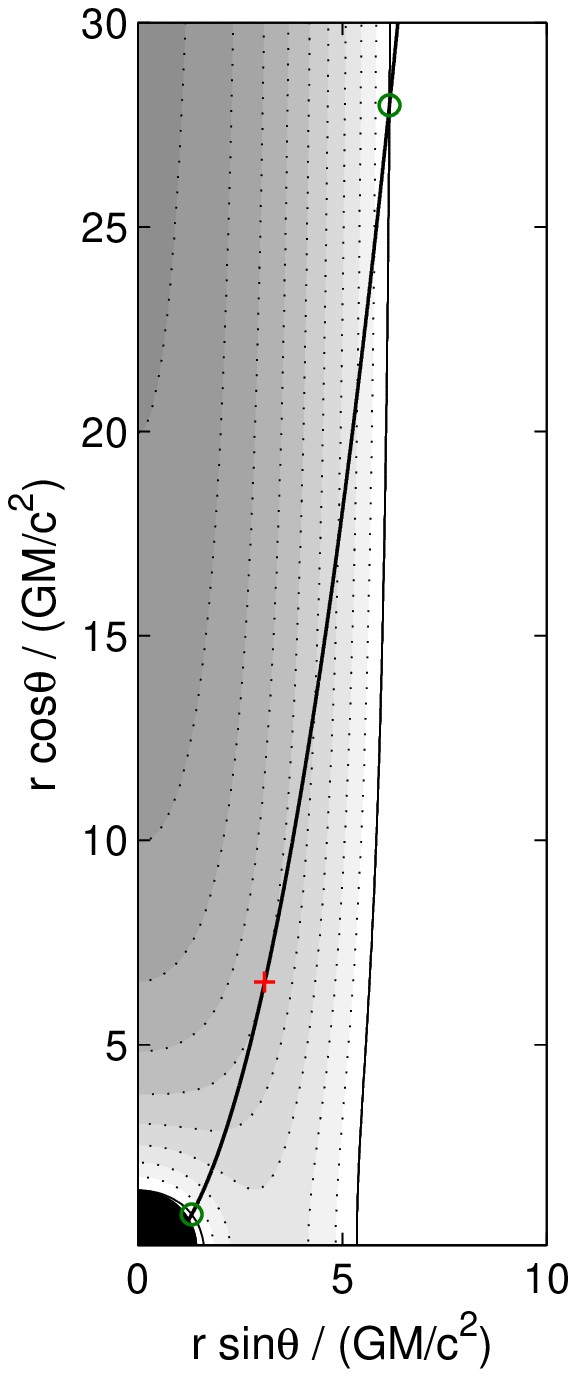}\\
\includegraphics[scale=0.38,angle=-90,trim=0cm 0.4cm 0cm 0cm, clip=true,width=0.4 \textwidth]{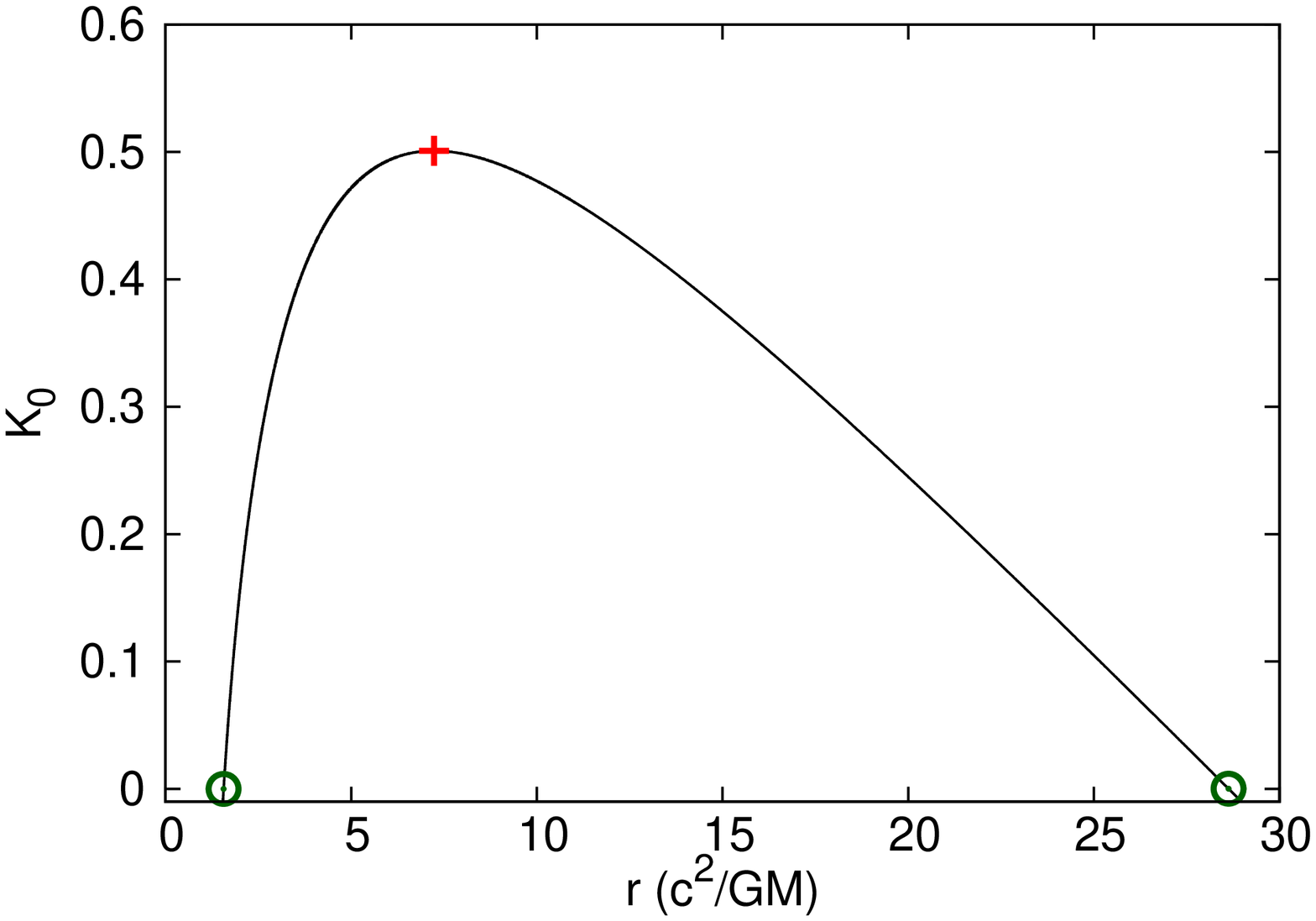}
\caption{Determining the location of the light surfaces and the separation surface for a field line. 
{\em Top}: 
Contour plot of $K_{0}$ for the case  $a=0.9$ and $\Omega_{\rm F}=\Omega_{\rm H}/2$. 
The black area represents the region enclosed by the black hole event horizon. 
The contour of $K_{0}=0$ is indicated by the solid line and
the contours in region $K_{0}>0$ are indicated by the dotted lines, with values at $K_{0}=0.1$, 0.2, \ldots, 0.8, 0.9. 
The thick solid line is 
a representative large-scale, black hole-threading field line.
The two green circles on it indicate the locations of the light surfaces, 
  which corresponds to $K_{0}=0$ 
  and the red cross indicates the location of the the separation surface 
at which $K'_{0}=0$. 
{\em Bottom}: 
The value of $K_{0}$ along the field line.  
The locations of the light surfaces and separation surface are indicated by the same symbols.}  
\label{fig:less}
\end{center}
\end{figure}

The separation surface is inside the region bounded between the two light surfaces, and is determined via searching for where $K_{0}^{'}=0$ along each flow streamline in the calculations.   
Figure \ref{fig:less} shows  
  how $r_{\rm s}$ on a specific field line (flow streamline) is determined 
  in the demonstrative case with $K_{0}(a,\Omega_{\rm F})=K_{0}(0.9,\Omega_{\rm H}/2)$,  
  (where $\Omega_{\rm H}$ is the angular velocity of the black hole 
  and $r_{+}$  is the radius of the outer event horizon). 
The location  where $K_{0}=0$ and $K'_{0}=0$ along the field line can be read from the 
    contours of $K_{0}$, which are part of the light surfaces and the separation surface, respectively.
Note that the locations of the light surfaces and the separation surfaces  
  are independent of the flow parameters, such as the mass loading,    
  as they are determined only by $K_{0}(a,\Omega_{\rm F})$ and its derivative, $K_{0}'$.  
 

\subsection{Critical Surfaces for cold GRMHD flows}
\label{sec:critical surface} 
Critical points appear when $D$ vanishes in the expression of $(\ln u_{\rm p})'=N/D$. 
In the cold limit, there are two critical points.
The Alfv\'en critical point corresponds to 
  where $u_{\rm p}$ is equal to the poloidal Alfv\'en speed, i.e.\ 
\begin{equation}
\label{eq:u2_aw}
    u^{2}_{\rm AW}(r;\Psi)=\frac{B_{\rm p}^{2}}{4\pi\mu n}K_{0}\; ,   
\end{equation} 
  and the fast magnetosonic critical point corresponds to where  
  $u_{\rm p}$ equals the fast magnetosonic speed, i.e.\  
\begin{equation}
\label{eq:u2_fm}
   u^{2}_{\rm FM}(r;\Psi)=u^{2}_{\rm AW}+\frac{B^{2}_{\phi}}{4\pi\mu n \Delta \sin^{2}\theta} 
\end{equation}
   \citep[see][]{T90}.

At the Alfv\'en surface  
\begin{equation}\label{eq:ma2a}
  M_{\rm A}^{2}=K_{0}~ \big\vert_{r_{\rm A}} \ . 
\end{equation}
Setting $u^{2}_{\rm p}=u^{2}_{\rm AW}$ yields  
\begin{equation}
   \frac{n}{4\pi\mu}K_{0}=\eta^{2}   \ .
\end{equation}
Since $n$ and $\eta$ are positive, $K_{0}>0$ at the Alfv\'en surface.    
The Alfv\'en surfaces are therefore 
constrained inside the region bounded by the light surfaces (where $K_{0}=0$).    
In addition, $K_{0}\to0$ as $\eta\to0$, 
  implying that the Alfv\'en surfaces approach the light surfaces when mass loading decreases.


Because the flow must be super-Alfv\'enic outside the light surfaces (when shocks are absent), would the flows downstream, outside of the light surfaces eventually reach fast magneto-sonic speeds? 
The answer to the above question is different for inflows and outflows.
For the inflow, the magneto-sonic speed is certainly reached, 
  as causality requires that the flow speed must surpass all the possible characteristic speeds  
  before the flow would enter the black hole event horizon \citep[][]{T90}. 
For the outflow, whether or not the flow speed will reach the fast magneto-sonic speed 
  depends on how fast the field decays along the flow \citep[][]{T98}.
 
 If the fast surface exists, 
the physical flow solution for the WE can be uniquely determined after specifying three of the conserved quantities, 
and searching for the last one until the the flow can smoothly pass the fast surface. 
(see, e.g. Appendix C in \citet[][]{pu12} for the case of inflow as a demonstration).
At the fast surface, where $u^{2}_{\rm p}=u^{2}_{\rm FM}$, we have
\begin{equation}
\label{eq:dis_fm}
    \frac{n}{4\pi\mu}\left(K_{0}+\frac{B^{2}_{\phi}}{B^{2}_{\rm p}\Delta \sin^{2}\theta}\right)=\eta^{2}\;.
\end{equation} 
By equation (\ref{eq:eta_def}), while all else being equal, a relatively smaller $\eta$ is expected to produce a stronger $B_{\phi}$ ($\propto F_{r\theta}$) (see \citet[][]{pud06} for a Newtonian version of such MHD feature). As a result, a smaller $K_{0}$ is required to satisfy equation (\ref{eq:dis_fm}) when a smaller mass loading is applied. That is, the location of the fast surface moves farther away from the light surface as the mass loading decreases. For a GRMHD inflow, the location of the fast surface gets more and more closer to the event horizon.

\end{document}